# Magnon confinement in a nanomagnonic waveguide by a magnetic Moiré superlattice

Jilei Chen,[1,2] Marco Madami,[3] Gianluca Gubbiotti,[4] and Haiming Yu[2,5]

[1] Shenzhen Institute for Quantum Science and Engineering, Southern University of Science and Technology, Shenzhen, 518055, China.

[2] International Quantum Academy, Shenzhen, 518048, China.

[3] Dipartimento di Fisica e Geologia, Università di Perugia, Perugia I-06123, Italy

[4] Cnr-Istituto Officina dei Materiali, Unità di Perugia, Via A. Pascoli, 06123, Italy

[5] Fert Beijing Institute, MIIT Key Laboratory of Spintronics, School of Integrated Circuit Science and Engineering, Beihang University, Beijing 100191, China

Authors to whom correspondence should be addressed:

chenjl6@sustech.edu.cn, gubbiotti@iom.cnr.it, and haiming.yu@buaa.edu.cn

# ABSTRACT

The study of moiré superlattices has revealed intriguing phenomena in electronic systems, including unconventional superconductivity and ferromagnetism observed in magic-angle bilayer graphene. This approach has recently been adapted to the field of magnonics. In this Letter, we investigate the confinement of spin waves in a nanomagnonic waveguide integrated on top of a magnetic moiré superlattice. Our numerical analysis reveals a magnonic flat-band at the centre of the Brillouin zone, created by a 3.5 degrees twist in the moiré superlattice. The flat-band, characterized by a high magnon density of states and a zero group velocity, allows for the confinement of magnons within the *AB* stacking region. The flat-band results from the mode anticrossing of several different magnon bands, covering a wavevector range of nearly 40 rad/µm and a 166 nm wide spatial distribution of the magnon trapping in the waveguide. Our results pave the way for nanomagnonic devices and circuits based on spin-wave trapping in magnon waveguides.

When two periodic two-dimensional layers are combined with a twist angle, they form a moiré superlattice, which exhibits different behaviours to the original single layer. In the field of two-dimensional materials, stacking two graphene layers at a twist angle of 1.1 degrees, known as the "magic angle", results in a Moiré superlattice. At the magic angle, the Fermi velocity goes to zero and the energy bands near the charge neutrality point (where upper and lower Dirac cones meet) become flat. Consequently, a high density of states is observed at this frequency, leading to remarkable electronic properties such as superconductivity, weak ferromagnetism and correlated insulator states.[1-7]

More recently, moiré physics has been integrated into the field of magnonics[8-12], which exploits the involvement of magnons (quasi-particles associated with spin waves in magnetic systems) to carry, manipulate and process information[13-19]. Spin waves can propagate in magnetic materials without the movement of electron charges, thus providing a very promising platform for making electronic devices without heat dissipation. Previously, the formation of a magnon flat band and the resulting strong two-dimensional magnon confinement in twisted bilayer magnonic crystals at the optimal "magic angle" has been reported [20]. In addition, we have experimentally observed spin-wave edge and cavity modes in twisted moiré superlattices consisting of two antidot lattices with circular holes etched into one and the same yttrium iron garnet thin film[21].

The study of spin wave propagation in nanoscale magnonic waveguides is essential for the design of magnonic devices and nanocircuits[22-25], such as magnonic directional couplers for logic computing[26]. Additionally, magnonic band gaps[27,28] can be formed at specific frequencies using one- and two-dimensional magnonic crystals[29-33], providing a method of control magnons in magnonic waveguides. The exceptional characteristics of moiré superlattices, including their tunable band structures and topological nontrivial transport properties, hold the potential for functional magnonic devices that have not been observed in conventional ferromagnetic materials.

In this Letter, we numerically investigate the confinement of spin waves in a nano-magnonic waveguide patterned atop a magnetic moiré superlattice. A flat magnonic band appears at the centre of the Brillouin zone ($k$=0) of two antidot magnonic crystal layers[34] which are twisted at an angle of 3.5 degrees. The appearance of the flat magnon band with zero group velocity leads to a high magnon density state, resulting in strong magnon confinement within the $AB$ stacking region. The flat band formation results from the mutual anticrossing of different magnon bands from the two magnonic crystal layers, facilitated by the magnon-magnon coupling between the layers. The spatial distribution of the magnon trapping exhibits a bandwidth of 166 nm in the nanomagnonic waveguide. This confinement is subsequently imprinted on the overlying magnonic waveguide as a consequence of the interlayer exchange interactions. The trapping spin waves in a magnonic waveguide offers great potential for the creation of functional magnonic devices and circuits.

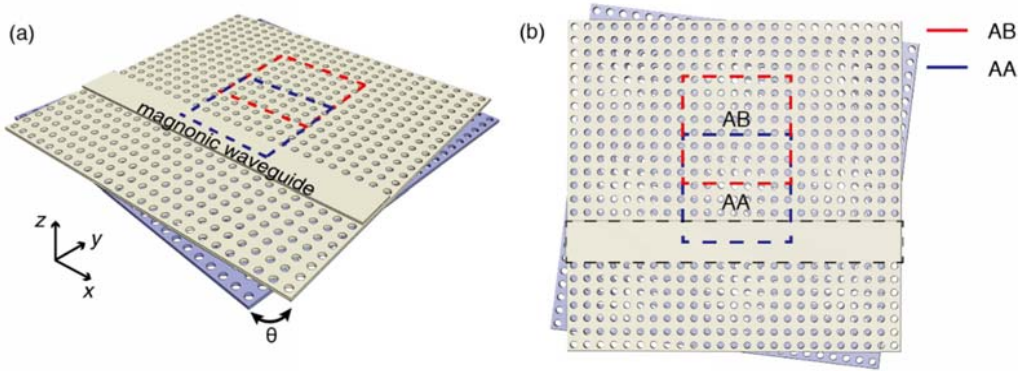

**FIG. 1.** (a) Side view of the magnetic Moiré superlattice based heterostructure. Two YIG magnetic layers with antidot lattices are twisted with an angle of $\theta = 3.5°$. The dimensions of the simulated area is 6 μm × 6 μm × 8 nm. The red and blues dashed squared regions indicate the $AB$ and $AA$ stacking regions of the moiré superlattices. A magnonic waveguide, consisting of a YIG stripe 500 nm wide, is patterned on top of the antidot lattice at the center of the AB

region. The external field is applied in the $y$ direction. (b) Top view of the magnetic Moiré superlattice based heterostructure. The red dashed (blue dashed) square represents the Moiré unit cell with *AB* (*AA*) stacking region at its center. The black dashed square represents the position of the nanomagnonic waveguide.

Schematic representations of the investigated magnetic moiré superlattice-based heterostructures under investigation are shown in Figures 1a and 1b. In the numerical simulations, ferrimagnetic yttrium iron garnet (YIG) layers with low damping properties are used. Both YIG layers are structured into antidot square lattices as the magnonic crystals. The YIG antidot lattices have squared symmetry with periodicity of $a_0$=100 nm and circular holes with diameter of 50 nm. The thickness of the YIG layers and the waveguide is 2 nm. The separation distance between two moiré layers is 2 nm. The saturation magnetization and magnetic damping are set as $M_S$ = 140 kA/m, and $\alpha$ = 0.0001, respectively. The interlayer exchange coupling is taken as free parameter to control the magnonic band structure and the flat band formation. The formation of the moiré superlattice is achieved by rotating one layer with respect to the other. In the simulation, we have set the rotation angle to ($\theta = 3.5°$), which corresponds to the previously reported 'magic angle' for magnonic nanocavity modes[20]. The period of the moiré superlattice can be calculated as $a_m = a_0/\theta$, where $a_0$ is the intrinsic period of the antidot square lattice. With a rotating angle of $\theta = 3.5°$, the dimensions of the moiré unit cell are 1.6 μm × 1.6 μm, as illustrated in Fig. 1b for both *AA* and *AB* stacking regions. In addition, a 500 nm wide magnonic waveguide is patterned on top of the Moiré superlattice as sketched in Fig. 1b. The waveguide is in direct contact with the moiré superlattice. For the simplicity, the waveguide material is also chosen to be YIG, with the exchange coupling between the waveguide and the topmost layer of the moiré superlattice kept at the bulk value of $A = 3.7$ pJ/m.

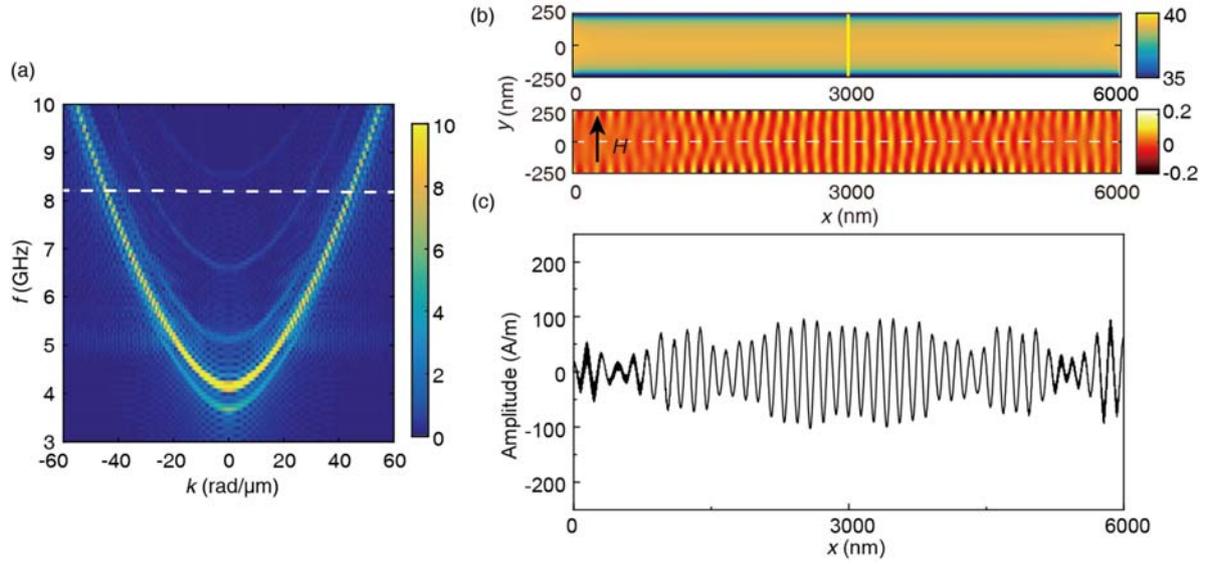

**FIG. 2.** (a) The calculated spin-wave band structure of the single magnonic waveguide. (b) The upper panel illustrates the internal field within the magnonic waveguide (unit in kA/m). The lower panel depicts the spatial distribution of the magnetization dynamics (unit in kA/m) $m_x$ at $t$=4 ns with the excitation frequency of 8.2 GHz, which is the same frequency as the flat band observed for the moiré superlattice shown in Figure 4a. The excitation region (yellow bar) is positioned in the middle of the simulated area. (c) is a linecut of the spatial map at $y$=0 nm.

First, we study the magnon propagation in a single magnonic waveguide with the width of 500 nm. The external magnetic field is applied in the $y$-direction with the field of 50 mT, which is related to the Damon-Eshbach spin-wave mode. The magnetic ground state, which is not uniform due to the demagnetization field, is first determined. A sinc pulse as the excitation field is applied in the middle of the magnonic waveguide (see Fig. 2 (b)), within the spatial range of 2.99 μm ⩽ $x$ ⩽ 3.01 μm. Therefore, the spin-wave wavevector $k$ is also along the $x$-direction. The calculated magnonic band structure for a single magnonic waveguide is depicted in Figure 2a. It exhibits the typical quadratic spin-wave dispersion, accompanied by several higher-order spin-wave modes associated with standing spin-wave width modes. The two lower modes are attributed to the spin-wave center and edge modes of the waveguide, due to the different internal fields shown in Fig.

2b. The spatial spin-wave distribution for magnon propagation in the magnonic waveguide is shown in Fig. 2b, with the excitation frequency of 8.2 GHz, corresponding to the spin-wave wavelength of 144 nm calculated from the dispersion relation. It can be observed that the spin waves can propagate in the single magnonic waveguide without any confinement. Fig. 2c displays a line scan of the spin-wave propagation profile at 8.2 GHz, extracted from the center of the magnonic waveguide (white dashed line in Fig. 2 (b)).

The precise control of spin-wave properties within magnonic waveguides is important for the realization of functional magnonic devices and circuits[35,36]. Numerous prior investigations have centered on diverse strategies to achieve this, such as the voltage control[37] and harnessing its nonlinearity[26]. In this work, we introduce the method of using artificial moiré superlattices. While recent theoretical works have highlighted intriguing magnonic properties in twisted bilayer magnets, particularly in the presence of Dzyaloshinskii-Moriya interactions[38] and stacking domain walls[39], these magnons operate at exceedingly high frequencies, which is challenging for practical implementation in coherent magnonics operating in the GHz range. Our investigation into artificial moiré superlattices, therefore, presents a promising approach for advancing microwave-based magnonics.

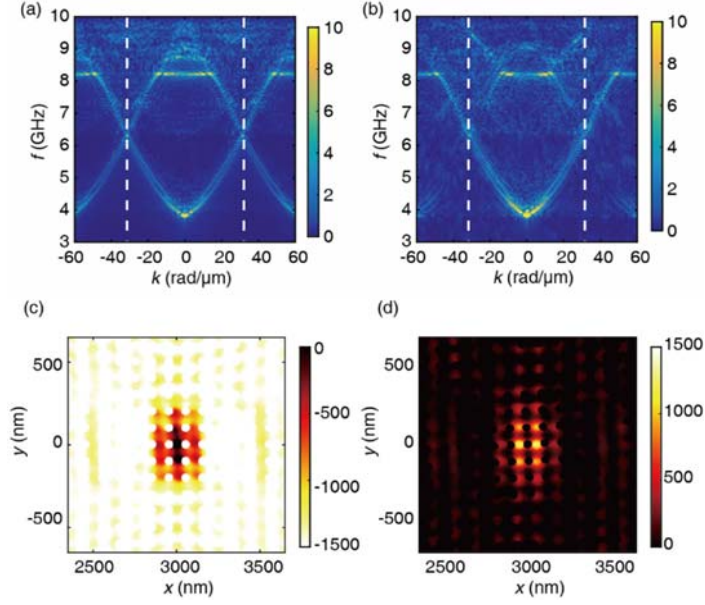

**FIG. 3.** (a) Spin-wave band structure of the bottom YIG layer of the twisted Moiré layers. (b) Spin-wave band structure of the top YIG layer. (c) The spatial distribution of the magnetization dynamics of the bottom YIG layer at $t = 4$ ns with the single frequency excitation of 8.2 GHz. (d) The spatial distribution of the magnetization dynamics of the top YIG layer.

We then investigate the formation of magnon flat bands in two twisted YIG layers with antidot lattices. The single magnonic crystal layer with antidot lattices behaves regularly and exhibits magnonic band gaps at Brouillon zone boundaries as shown in Ref. 20. By twisting one YIG layer with respect to the other, the interlayer exchange coupling hybridizes different magnonic branches in the band structure through the interlayer magnon-magnon coupling. This results in an anticrossing behavior at the meeting points of the magnon bands from the two YIG layers. Finally, the magnon mode hybridization evolves into a quasi-flat band at a certain twist angle and interlayer exchange coupling, comparable to the formation of flat bands in magic-angle bilayer graphene systems[3,4]. The effective interlayer exchange field is expressed as $\mathbf{h}_{\text{eff}} = A_{12}|\nabla_z \mathbf{m}|^2$, where the $z$-direction magnetization gradient corresponds to the magnon-magnon coupling induced by the interlayer exchange interaction. At the center of the *AB* region, the twisted magnetization distribution difference in the *z*-direction between two adjacent YIG layers is maximized, resulting in the greatest magnon-magnon hybridization of the magnon bands. This eventually leads to the formation of the magnon flat band. Figures 3a and 3b show the spin-wave band structure of the two twisted YIG magnonic crystal layers. Note that the nanomagnonic waveguide is already considered on top of the moiré superlattices. Therefore, the interlayer exchange coupling between two layers is set to $A_{12} = 15$ µJ/m$^2$, which is not the value for the flat band in the system with only two twisted YIG layers[20]. The spatial distribution of the magnetization dynamics, denoted as $m_x$, in the confined spin waves within the *AB* region of the Moiré superlattices is presented in Figures 3c and 3d. These correspond to the flat band frequency of 8.2 GHz. The confined spin waves in the two layers form nanomagnonic cavities, where they exhibit out-of-phase oscillation precession due to the twist angle. Note that the physical mechanism of the spin-wave confinement is fundamentally different from the forbidden band in a magnonic crystal, which relies on the magnonic band gap. The spin-wave confinement resulting from the flat band observed at the center of the BZ (*k*=0) is a consequence of the large magnon density of states.

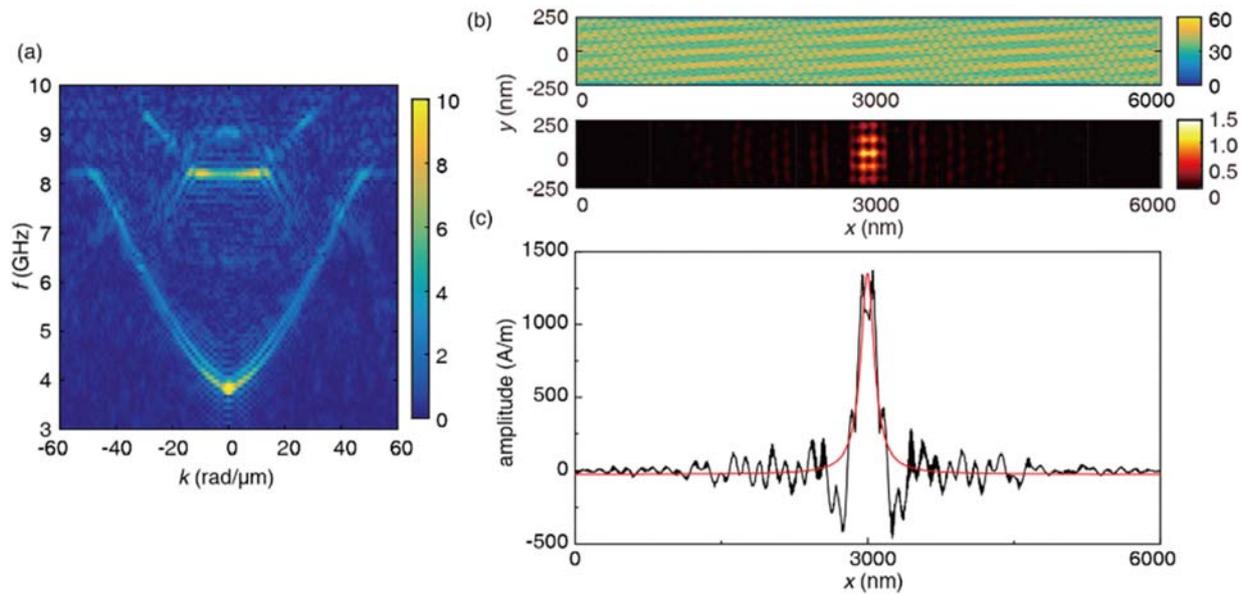

**FIG. 4.** (a) Magnonic bandstructure of the magnonic waveguide on top of the Moiré superlattice. (b) The upper panel illustrates the internal field within the magnonic waveguide (unit in kA/m). The lower panel depicts the spatial distribution of the magnetization dynamics (unit in kA/m) $m_x$ at $t = 4$ ns. (c) A line-cut of the magnetization amplitude at $y=0$ in (b). The red curve is a Lorenz fitting which yields a linewidth of 166 nm.

Figure 4a shows the simulated spin-wave band structure within the magnonic waveguide patterned over the *AB* region of the Moiré superlattice, where a flat band appears at approximately 8.2 GHz, indicating the zero spin-wave group velocity and the formation of the magnon trapping. The flat band spans a wave vector range from -19.6 rad/μm to 19.6 rad/μm. Due to the large demagnetization field of the antidot lattice, the internal field in the waveguide shown in Fig. 4b is completely different from Fig. 2b. The spatial spin-wave configuration for the magnon confinement within the magnonic waveguide is shown in Fig. 4b, where the excitation occurs at the flat band frequency of 8.2 GHz, which is not consistent with the observation of the flat band in the Figs. 3a and 3b. The flat band behavior of the moiré superlattice is imprinted into the magnonic waveguide topmost layer by the exchange interaction at the interface. A single spin-wave spectrum extracted from the center of the magnonic waveguide is displayed in Fig. 4c. A Lorenzian fit of this spectrum reveals a linewidth of 166 nm, which approximates the wavevector broadband of $\Delta k = \frac{2\pi}{\Delta x} = 37.9$ rad/μm. In contrast to a resonator coupled to a waveguide[40,41], the moiré superlattice, with its flat-band property, not only amplifies the magnon population but also confines these magnons within a nanoscale regime, giving rise to a nanomagnonic cavity. The magnon amplification is an important mechanism to counteract natural damping[42] in magnetic materials, thereby sustaining or enhancing the intensity of spin waves.

Oscillations of magnetizations can be observed beyond the magnon trapping region, attributed to the non-zero magnon group velocity at the edges of the flat band. It is important to note that spin-wave trapping occurs only when the magnonic waveguide is patterned over the *AB* stacking region. Conversely, as demonstrated in the supplementary material, spin waves can propagate from the stacking region in the *AA* stacking configuration. The realization of spin wave confinement in a magnonic waveguide could be a crucial step toward achieving a magnon condensation for reservoir purposes.

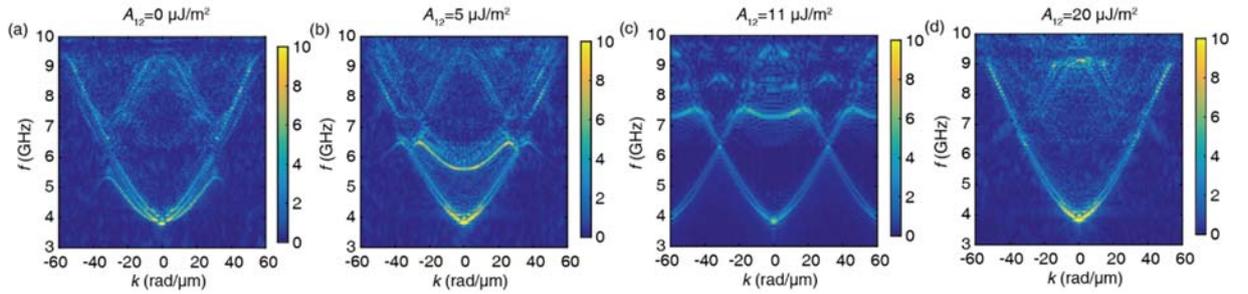

**FIG. 5.** Moiré flat-band of the magnonic waveguide with different interlayer exchange coupling. (a)-(d) Simulated spin-wave band structure of the magnonic wavegude at the Moiré unit cell center (*AB* stacking region) with interlayer exchange coupling of (a) 0 μJ/m$^2$, (b) 5 μJ/m$^2$, (c) 11 μJ/m$^2$ and (d) 20 μJ/m$^2$.

In the following, we investigate the formation of a moiré flat band in a magnonic waveguide by adjusting the interlayer exchange coupling between two adjacent YIG layers. When the interlayer exchange coupling is zero and only the interlayer dipolar coupling exists, the flat band cannot form in the magnonic waveguide, and the spin-wave band structure behaves like a conventional antidot magnonic crystal, as shown in Fig. 5a. This indicates that the interlayer exchange coupling is essential for the formation of the magnonic flat band in the magnonic waveguide. When we increase the interlayer exchange coupling to $A_{12} = 5$ μJ/m$^2$, a curved magnon band around 6 GHz is formed, which is due to the hybridization of magnon modes in different YIG layers. However, this curved band does not support confined spin-wave modes due to the low magnon density of states at a single frequency. If the interlayer exchange coupling is further increased to $A_{12} =$ 11 μJ/m², the curved band shifts up to around 7 GHz and becomes flatter. Finally, at $A_{12} =$ 15 μJ/m², as shown in Fig. 4a, the curved band becomes completely flat, allowing the spin wave amplitude to be confine spin waves within the magnonic waveguide. Notably, spin waves can still propagate freely in the magnonic waveguide at frequencies beyond the flat band frequency, providing flexibility for controllable magnonic devices. When we further increase the interlayer exchange coupling to $A_{12} = 20$ μJ/m², the Moiré flat band shifts up to approximately 9 GHz with distorted behaviour, and the wavevector range becomes more restricted. The moiré flat band sustains a high-quality flatness with the interlayer exchange coupling $A_{12} = 15$ μJ/m². Such broad wavevector range enables an accumulation of magnon intensity inside the nano-waveguide, which provides perspectives to further realization of magnonic applications, such as magnon Bose-Einstin condensation[43] and magnon laser[44].

This work introduces a method for confining magnons via a magnetic Moiré superlattice, as demonstrated through micromagnetic simulations. In terms of the experimental realization, the advanced cutting-edge nanofabrication technologies would be helpful. Specifically, low-damping YIG thin films can be grown through sputtering[45,46], while antidot lattices can be precisely crafted using e-beam lithography and ion-beam etching[47-49]. Furthermore, for the spatial characterization of magnon confinement, techniques such as micro-focused Brillouin light scattering[21,26] or nitrogen-vacancy (NV) magnetometry[50] offer powerful mapping capabilities. In addition, recent experimental investigations into magnon propagation in synthetic antiferromagnets[51,52] have illuminated potential routes for experimentally realizing the proposed heterostructure by using magnetic moiré superlattices. However, the fabrication of layered antidot structures is challenging because the twist angle and patterning of multiple layers has to be precisely controlled. Moreover, the interlayer exchange coupling is required to be fine-tuned, as fabrication imperfections could significantly impact its precision, thereby affecting the overall performance of the heterostructure.

In conclusion, we numerically calculate the magnonic band structure of a nanomagnonic waveguide patterned on top of a magnetic Moiré superlattice. A flat band at a single frequency emerges when the waveguide is atop the *AB* stacking region. The high magnonic density of states at the flat band results in magnon confinement within the waveguide, covering around 166 nm in the spatial range and around 40 rad/µm in the wavevector range. Spin waves are free to propagate outside the flat band frequency. This work introduces a promising approach to control spin waves in nanomagnonic waveguides, which may further be utilized to design nanomagnonic architectures. Moreover, future research could explore optimizing the waveguide dimensions for specific applications, investigating the impact of the twist angle, external fields on the magnonic properties, and exploring potential applications in nanomagnonic devices and circuits.

See the supplementary material for the spin waves propagation from the region in the *AA* stacking configuration.

We wish to acknowledge the support by NSF China under Grants No. 12104208 and No. 12074026, the Guangdong Basic and Applied Basic Research Foundation (Grant No. 2023A1515011622) and the National Key Research and Development Program of China Grant No. 2022YFA1402801.

G. G. acknowledges the European Union—Next Generation EU under the Italian Ministry of University and Research (MUR) National Innovation Ecosystem Grant No. ECS00000041—VITALITY-CUP B43C22000470005. G.G. also acknowledges funding from the European Union – Next Generation EU – "PNRR – M4C2, investimento 1.1 – "Fondo PRIN 2022" –TEEPHANY–ThreEE-dimensional Processing TecHnique of mAgNetic crYstals for magnonics and nanomagnetism ID 2022P4485M- CUP D53D23001400001".

M. M. acknowledges financial support from the Italian Ministry of University and Research through the PRIN-2022 project entitled "Metrology for spintronics: A machine learning approach for the reliable determination of the Dzyaloshinskii-Moriya interaction (MetroSpin)", cod. 2022SAYARY-CUP J53D23001470006.

**AUTHOR DECLARATIONS**

**Conflict of Interest**

The authors have no conflicts to disclose.

**Author Contributions**

Jilei Chen: Investigation (equal); Methodology (equal); Writing – original draft (equal). Marco Madami: Conceptualization (supporting); Supervision (supporting); Writing – review & editing (supporting). Gianluca Gubbiotti: Conceptualization (equal); Funding acquisition (equal); Project administration (equal); Supervision (equal); Writing – review& editing (equal). Haiming Yu: Conceptualization (equal); Funding acquisition (equal); Supervision (equal); Writing – review& editing (equal).

**DATA AVAILABILITY**

The data that support the findings of this study are available from the corresponding author upon reasonable request.